\documentclass[preprint]{elsarticle}
\usepackage{hyperref,color}
\usepackage[displaymath, mathlines]{lineno}
\modulolinenumbers[1]
\usepackage{todonotes}
\usepackage{ulem}
\usepackage{amssymb}
\usepackage{booktabs}
\usepackage{chemformula}
\usepackage{gensymb}
\usepackage{comment}
\usepackage{svg}
\setchemformula{radical-radius=1pt}
\RequirePackage{xspace}
\DeclareFontFamily{OT1}{pzc}{}
\DeclareFontShape{OT1}{pzc}{m}{it}{<-> s * [1.200] pzcmi7t}{}
\DeclareMathAlphabet{\mathpzc}{OT1}{pzc}{m}{it}
\usepackage{amsfonts}
\usepackage{textcomp}
\journal{Nuclear Instruments and Methods A}

\usepackage{multirow}
\newcommand{\sipm}{SiPM}
\newcommand{\sipms}{SiPMs}

\newcommand{\cosixty}{$^{60}\mathrm{Co}$} 
\newcommand{\doserate}{\ensuremath{\mathcal{R}}}
\newcommand{\gdoserateh}{1.3}
\newcommand{\gdoseratel}{0.3}
\newcommand{\gdoseratewarm}{1.6}
\newcommand{\tyvek}{Tyvek\textsuperscript{\textregistered}}
\newcommand{\kapton}{Kapton\textsuperscript{\textregistered}}
\newcommand{\keV}{\ensuremath{\,\text{ke\hspace{-.08em}V}}\xspace}

\begin{document}

\begin{frontmatter}

\title{Reduction of light output  of plastic scintillator tiles during irradiation  at cold temperatures and in low-oxygen environments}

\author[umd]{B.~Kronheim\corref{cor}}
\ead{bkronhei@umd.edu}
\author[umd]{A.~Belloni}
\author[umd]{T.K.~Edberg}
\author[umd]{S.C.~Eno}
\author[umd]{C.~Howe}
\author[umd]{C.~Palmer}
\author[umd]{C.~Papageorgakis}
\author[umd]{M.~Paranjpe}
\author[umd]{S.~Sriram}

\cortext[cor]{Corresponding author}

\affiliation[umd]{ 
    organization={Dept. Physics, U. Maryland},
    city={College Park},
    state={MD},
    country={USA}
}

\begin{abstract}
The advent of the silicon photomultiplier has allowed the development of highly segmented calorimeters using plastic scintillator as the active media, with photodetectors embedded in the calorimeter, in dimples in the plastic.  To reduce the photodetector's dark current and radiation damage, the high granularity calorimeter designed for the high luminosity upgrade of the CMS detector at CERN's Large Hadron Collider will be operated at a temperature of about
-30\degree\,C.  Due to flammability considerations, a low oxygen environment is being considered.
However, the radiation damage to the plastic scintillator during irradiation in this operating environment needs to be considered.
In this paper, we present measurements of the relative decrease of light output during irradiation of small plastic scintillator tiles read out by silicon
photomultipliers.
The irradiations were performed using a \cosixty\ source both to produce the tiles' light
and as a source of ionizing irradiation
at dose rates of \gdoseratel , \gdoserateh , and \gdoseratewarm\,Gy/hr,
temperatures of -30, -15, -5, and 0\degree\,C,  and with several different oxygen
concentrations in the surrounding atmosphere.
The effect of the material used to wrap the tile was also studied.
Substantial temporary damage, which annealed
when the sample was warmed, was seen during the low-temperature irradiations, regardless of the oxygen concentration and wrapping material.  The relative light loss was largest with 
3M\textsuperscript{\tiny TM}
Enhanced Specular Reflector Film wrapping and smallest with no wrapping, although
due to the substantially higher light yield with
wrapping, the final light output is largest with wrapping.
The light loss was less at warmer temperatures.
Damage with 3\% oxygen was similar to that in standard atmosphere.   Evidence of a plateau in the radical density was seen for the 0\degree\,C data.
\end{abstract}


\end{frontmatter}

\section{Introduction}
\label{sec:intro}


Plastic scintillator is used in a wide variety of detectors for high energy physics~\cite{kharzheev} due to its low cost and high light output.  
It is, however, subject to radiation damage~\cite{dubna, Busjan199989}. 
During irradiation, chemical bonds in the long polymer molecules of the plastic are broken, producing highly reactive chemical radicals.  These radicals generally strongly absorb photons of the wavelengths typically produced by plastic scintillators, resulting in a reduced light output.  The radicals can re-bond to each other, to oxygen in the plastic, or to other parts of the molecule or other molecules (``cross linking"). Generally all of these mechanics produce molecules that are less absorptive than the free radicals.
Free radicals, before re-bonding, produce ``temporary" radiation damage.
After the end of irradiation, oxygen penetrates the plastic, re-bonding the remaining radicals.  The remaining absorptive centers (due to the oxides or cross linking) are referred to as ``permanent" damage.

The permanent damage after the end of irradiation and  after annealing has been well studied at room temperature and in standard atmosphere.   The damage depends on the dose rate, with larger damage per dose at lower dose rates~\cite{christos,Sirunyan_2020}. 
Permanent damage can be roughly parameterized using an exponential function

\begin{equation}
    \frac{L(d)}{L_0}=e^{-d/D}
    \label{eqn:doseconst}
\end{equation}
where $L(d)$ is the light output at a dose $d$, $L_0$ is the light output before irradiation, and $D$ is the ``dose constant," whose value depends on the type of scintillator used, its geometry, and the dose rate.  
However, studies of temporary damage are rare, due to the difficulty of such measurements. Measurements at room temperature include Refs.~\cite{Busjan199989,Sirunyan_2020, hgcaltdr}.
A literature search did not uncover any studies of temporary damage in scintillators at lower temperatures.

 The CMS experiment~\cite{CMS_ex} at CERN's Large Hadron Collider LHC in Geneva, Switzerland, will upgrade its endcap calorimeter in preparation for the high-luminosity upgrade of the LHC.  The new ``high granularity" calorimeter~\cite{hgcaltdr} (HGCAL) uses plastic scintillator in sections of the detector where lower hadronic shower intensity results in low integrated radiation deposition.  The entire calorimeter will be enclosed in a cryostat to maintain
a temperature of -30\degree\,C in order to reduce the dark current and the effects of radiation damage to the photodetectors 
used in the parts of the calorimeter instrumented with scintillator. This low temperature operation has the added benefit of shielding the portions of the detector instrumented with silicon sensors from similar radiation damage.
The oxygen content in the encased atmosphere may be limited to 3\% for safety reasons.  
When the CMS detector has recorded its entire expected luminosity (3000\,fb$^{-1}$), the anticipated integrated dose experienced by portions of the scintillator-instrumented regions of HGCAL can be as much as 2.5\,kGy. Peak radiation rates are expected to be as high as 0.15\,Gy/hr. 
Predictions of the permanent damage  for HGCAL were parameterized as a function of dose rate~\cite{hgcaltdr} as

\begin{equation}
    D=36~\mathrm{kGy} \sqrt{\frac{\doserate}{10~\mathrm{Gy/hr}} }
    \label{eqn:oldprediction}
\end{equation}
where \doserate\ is the dose rate and $D$ is the dose constant.
For a dose rate of 1.3\,Gy/hr, this corresponds to 
$D=13$~kGy.

In this paper, we report the results of measurements during irradiation of the 
light loss  of plastic scintillator tiles  due to temporary damage  relative to that before irradiation.
The tiles are in the form of  small rectangles with a spherical cap
dimple at the center of one face of each tile.
The light output was measured using a silicon photomultiplier (\sipm) placed
in the center of the dimple.
The  tiles are similar to those that will be used for the HGCAL.
The irradiation was done using a \cosixty\ source
in the Radiation Effects Facility at the Goddard Space Flight Center in Greenbelt, Maryland, at
dose rates of \gdoseratewarm, \gdoserateh, and \gdoseratel\,\,Gy/hr. 
The  \cosixty\ source was used both to produce the tiles' light signal
and as the source of ionizing irradiation.
Measurements were performed at temperatures
of -30, -15, -5, and 0\degree\,C, in atmospheres of 100\% nitrogen, 97\% nitrogen--3\% oxygen, and in dry air.
Tiles were studied encased in a variety of wrapping materials:  3M\textsuperscript{\tiny TM} Enhanced Specular Reflector Film (ESR), \tyvek, ESR punctured with openings to allow better gas diffusion, and
optical black paper wrapped in aluminum tape.

Reference~\cite{Busjan199989} presents a similar study for BCF-12 and  SCSF81(Y7)  wave-length-shifting fibers with polystyrene (PS) core and polymethyl methacrylate (PMMA) cladding, using a 100\,\keV X-ray source for 
dose rates of 7\textemdash 42\,Gy/h in argon and in air, and at room temperature.
In that paper, the authors were able to separately measure light production and absorption, while we only measure the combination for our tile geometry.  
A previous measurement of the permanent damage after irradiation  in standard
atmosphere at
a dose rate of 3\,Gy/hr using a \cosixty\ irradiator for plastic rods at -20\degree\,C and annealed in an environment
around -15 to -20\degree\,C
is given in Ref.~\cite{Ricci-Tam:2301237}.  That
measurement saw little sign of temporary damage.
Another set of measurements, done during irradiation
at room temperature, in standard atmosphere for tiles with embedded wavelength-shifting fiber, in the CMS detector at the LHC, is given in Ref.~\cite{Sirunyan_2020}. That paper saw a dose-rate effect and small temporary damage.
Reference.~\cite{9721821} presents studies of the temperature dependence of the refractive index and light yield  over a wide range of temperature.  They find that a 10\degree\,C temperature rise would correspond to a ~2\% light yield loss.

This paper is organized as follows.
First, a brief reminder of the theory of radiation damage in plastic scintillators is given in Section~\ref{sec:theory}.  
The materials and measuring circuit are described in Section~\ref{sec:setup}.
Details of the irradiations are described in Section~\ref{sec:irrad}.
The calibration of the \sipms\ is described in Section~\ref{sec:sipmcalib}.
The results are given in Section~\ref{sec:results}.
Finally, concluding remarks are given in Section~\ref{sec:conclusion}.

\section{Radiation damage of plastic scintillator}
\label{sec:theory}
The rate of radical production during irradiation goes as~\cite{Busjan199989}

\begin{equation}
\frac{d[Y]}{dt}=gQ\doserate-k[Y]^2 - s [Y][C_0]
\label{eqn:exp2}
\end{equation}
where $[Y]$ is the density of radicals, $g$ is the radiation-chemical yield, $Q$ is the scintillator density,  $k$ is the reaction constant for the re-bonding of a radical with another radical,
$s$ is the reaction constant for the bonding of a radical to oxygen, and $C_0$ is the oxygen concentration.
The first term represents the creation of radicals, the second term represents two unterminated radicals recombining to neutralize (``second-order" termination), and the third term represents the bonding of a radical to oxygen (``first-order" termination).
When $C_0$ is zero (no oxygen) and
at short times, when the second term is small compared to the first, integration yields a radical density that is proportional to the integrated dose $d$: $Y=gQd$. The second term grows with time; a steady state is eventually reached when the creation and termination of radicals are equal. In this case, the radical density becomes constant with time, and is no longer proportional to the integrated dose.   For a very simple propagation and termination chain with zero oxygen, the yield goes as the inverse of  the square root of the dose rate~\cite{SpinksWoods}.

Oxygen has a beneficial effect on radiation damage during irradiation, as it interacts strongly with radicals and terminates them,  preventing their buildup.
In the presence of oxygen, before irradiation, the surrounding oxygen diffuses into plastic.  The steady-state concentration is given by Henry's law

\begin{equation}
    C=SP
\end{equation}
where $C$ is the oxygen concentration in the sample, $S$ is the solubility and $P$ is the  oxygen partial pressure.
$S$ has an exponential (``van't Hoff-type") dependence on temperature~\cite{PPPH}

\begin{equation}
    S = S_0 e^{\frac{-\Delta H_s}{RT}}
\end{equation}
where $\Delta H_s$ is the molar heat (enthalpy), $R$ is the universal gas constant, and $T$
is the temperature in degrees Kelvin. 

During irradiation at room temperature, oxygen present in the plastic interacts strongly with radicals created during the irradiation.
Any initial oxygen present in the plastic is quickly consumed by bonding to radicals.  While it can be replenished
from outside, it is consumed by radicals, so that its penetration depth at steady-state is a function of dose rate.
The penetration depth  $z_0$ for oxygen into a rectangular slab of plastic is~\cite{cunliffe,seguchi}

\begin{equation}
z_0^2=\frac{2 \, M \, C_0}{\Upsilon \, \doserate}=\frac{2 \, M \, S \, P}{\Upsilon \, \doserate} = {\frac{\gamma}{\doserate}}
\label{eqn:z0}
\end{equation}
where \doserate\ is the dose rate,
$M$ is the diffusion coefficient for oxygen, $C_0$ is the oxygen concentration at the plastic's surface, $\Upsilon~(=gQ)$ is the specific rate constant of active site formation, $S$ is the oxygen solubility, and $P$ is the external oxygen partial pressure. 
As with $S$, $M$ has an exponential (``Arrhenius equation") dependence on temperature~\cite{vieth,difftemp,GILLEN1993101}.
In general, plastic scintillators whose thickness is greater than the oxygen penetration depth show the strong discoloration that is the signature of radicals at depths greater than the penetration depth.
That discoloration goes away (``anneals") at the end of irradiation as oxygen reenters the sample.  Plastic scintillators
thinner than the penetration depth of oxygen
that are irradiated at room temperature show little annealing, as the radicals are terminated as quickly as they are formed.

\section{Description of measurement apparatus}
\label{sec:setup}

The tiles used for our measurements are rectangular, with dimensions 
$3.4 \times 3.4 \times 0.3 ~{\rm cm}^3$. Each tile has
 a spherical cap indentation, called a ``dimple," at the center of one of the large faces, with a radius of
6.1\,mm and a depth of 1.6\,mm.  These dimples house the \sipm s.
Three different scintillator materials are used:  
cast polyvinyl toluene (PVT) based EJ-208 from Eljen Technology,\footnote{Eljen Technology, 1300 W. Broadway, Sweetwater, Texas 79556, United States} 
PVT-based BC-412 from Luxium Solutions\footnote{Luxium Solutions, 17900 Great Lakes Parkway, Hiram, Ohio, USA} (formerly Saint-Gobain), 
and
a polystyrene (PS) based molded tile produced at Fermi National Accelerator Laboratory\footnote{Fermi National Accelerator Laboratory Batavia IL 60510-5011} (FNAL). 
Some of the tiles were wrapped in a reflective coating, either \tyvek\ or ESR. All wrappings had holes aligned over the \sipm\ dimple.

\begin{figure}[hbtp]
\centering
\includegraphics[width=0.3\textwidth]{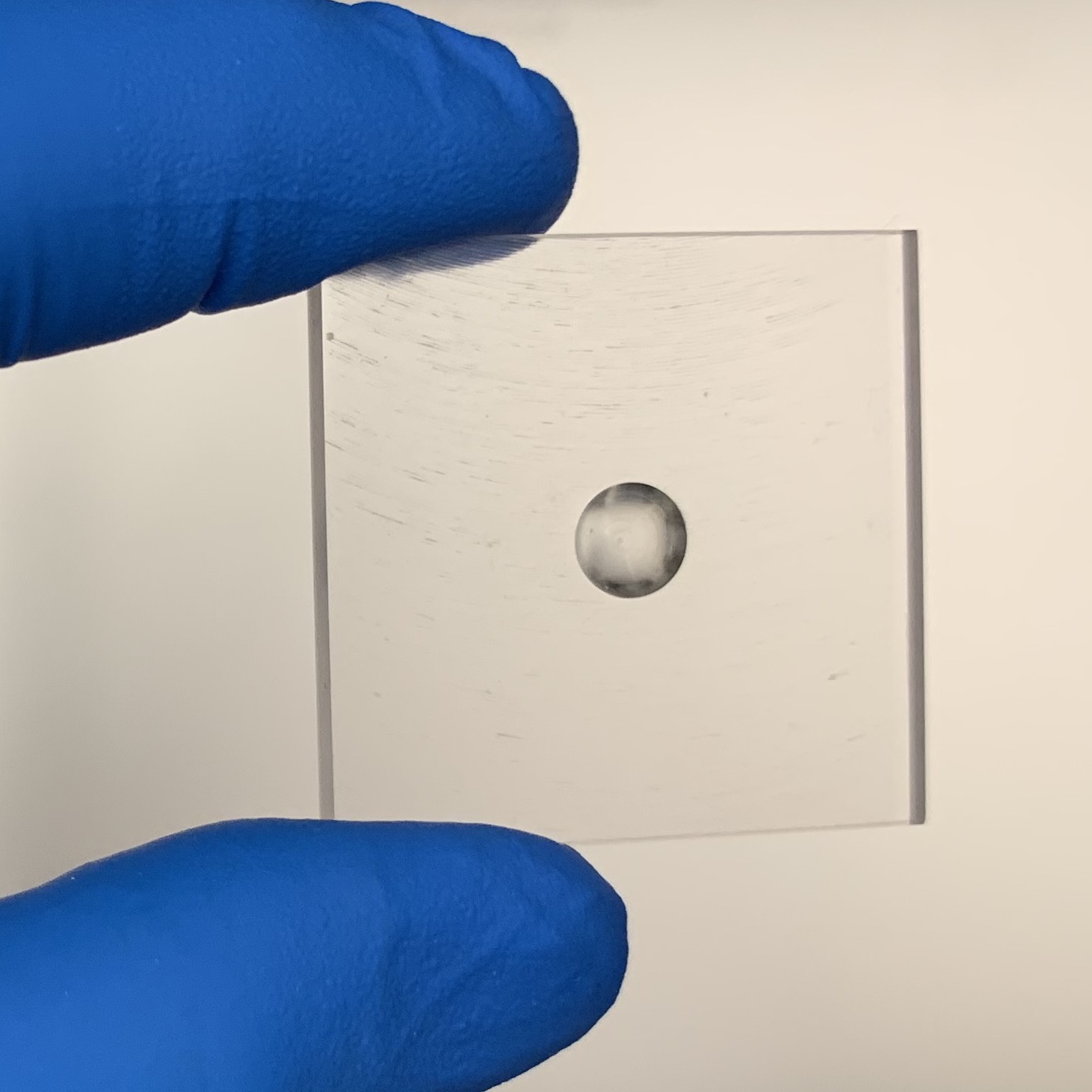}
\includegraphics[width=0.3\textwidth]{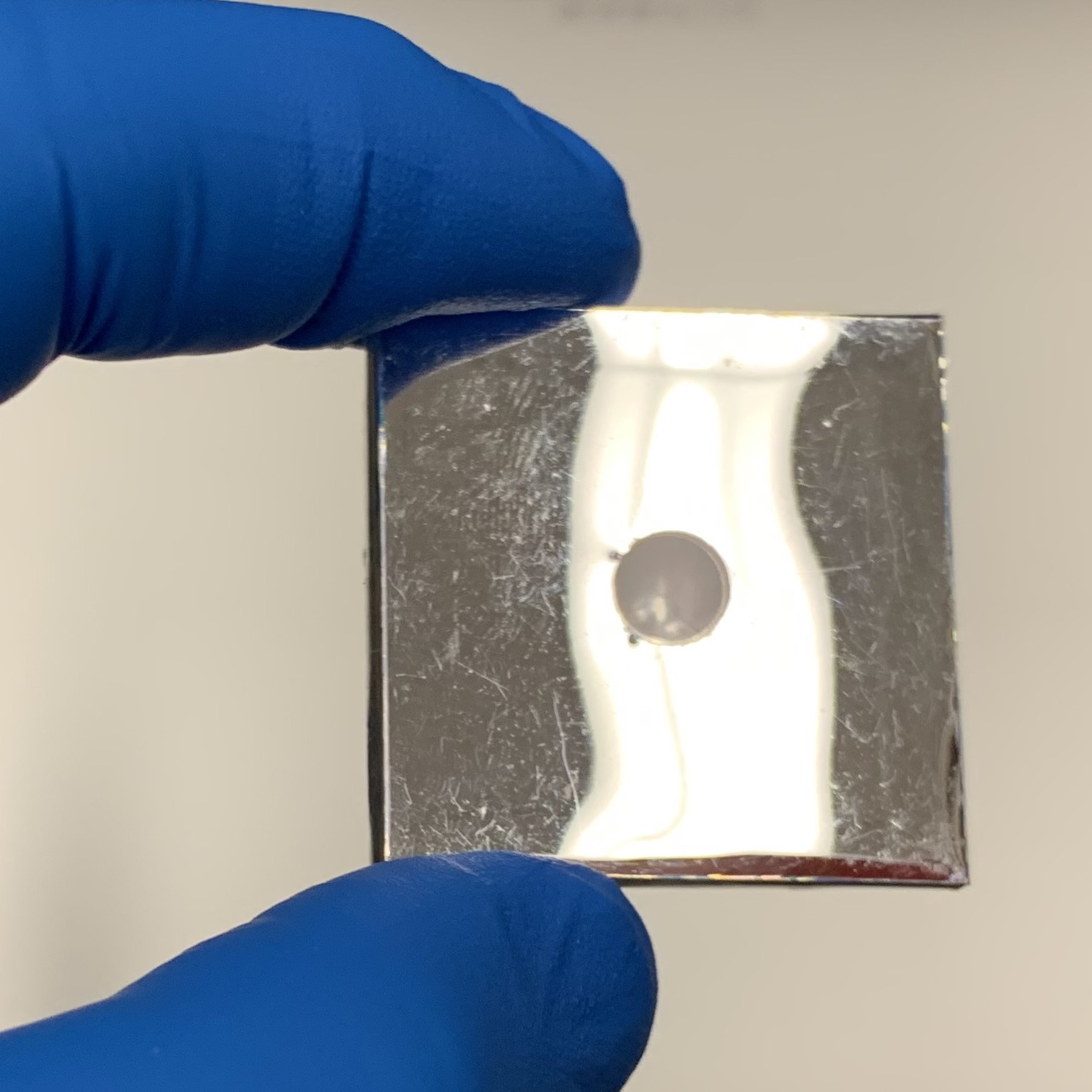}
\includegraphics[width=0.3\textwidth]{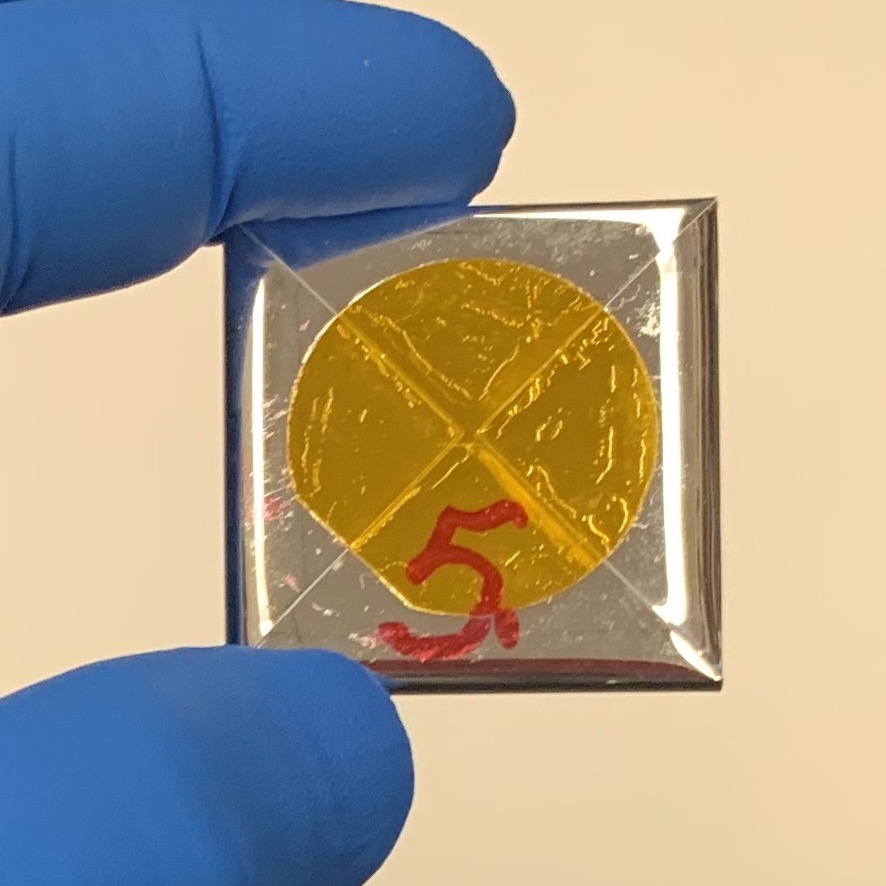}

\caption{
[Left] Photograph of a typical scintillating tile used in these measurements before wrapping.  This tile was made from EJ-208 scintillator;
[Center] A tile wrapped in ESR, showing the hole over the dimple. The apparent discoloration is due to reflections from the environment;
[Right] A tile wrapped in ESR, showing the wrapper closed with tape on the face opposite the dimple.
}
\label{fig:plastic}
\end{figure}

One tile was wrapped in black optical paper, covered by aluminum tape.
For one of the ESR wrappings, four 2.2\,mm diameter holes, each centered on an edge, approximately 5\,mm in from the edge, were added for improved gas access to the tile. For another new ESR wrapping,
razor blade slits, approximately 1.5\,cm long, centered on each edge, were added, also to improve gas access.
Wrappings were fastened with a sticker and also with \kapton\ tape.
ESR wrappings were shaped into a tight envelope around the tile, while the \tyvek\ wrappings were loose.
Figure~\ref{fig:plastic} [left] shows a typical tile without wrapping, Fig.~\ref{fig:plastic} [center] shows the hole in the  ESR wrapping over the dimple, and Fig.~\ref{fig:plastic} [right] shows the ESR closure on the face away from the dimple.

The \sipms\ were the S14160-1315PS \sipm\ from  Hamamatsu (HPK)~\footnote{Hamamatsu Photonics, 325-6, Sunayama-cho, Naka-ku, Hamamatsu City, Shizuoka Pref., 430-8587, Japan}, except that the glass window was replaced  by the manufacturer with epoxy for increased radiation hardness.
They have an active area of $1.4 \times 1.4\,{\rm mm}^2$.  

\begin{figure}[hbtp]
\centering
\includegraphics[width=0.7\textwidth]{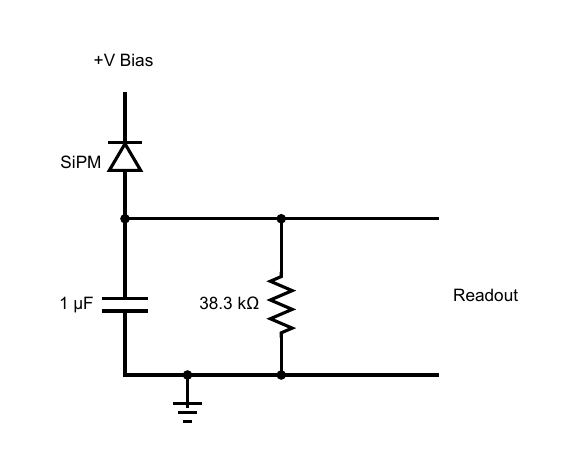}
\caption{
Circuit used to convert \sipm\ current to voltage. The nearly DC voltages were passed through 15\,m coaxial cables to the DAQ equipment outside the irradiation room.
}
\label{fig:pics}
\end{figure}

The currents from the \sipm\ were converted to a DC voltage using the RC circuit (38.3\,k$\Omega$, 1\,$\mu$F) shown in Fig.~\ref{fig:pics}.
The signal was read across the resistor.
 The passive components were tested for radiation damage to 881\,Gy; no damage was seen.

The \sipms\ and passive electronic components were soldered to circuit boards, located inside 2.25" x 2.25" x 4" or 2.25" x 2.25" x 5" aluminum boxes with 1\,mm thick walls (Bud \#CU-3003-A and \#CU-3004-A).  Tiles were affixed to the circuit board with \kapton\ tape. The boxes were sealed with several layers of aluminum tape to make them light-tight and to prevent air ingress.

Each box had dry gas flowing through it at $\sim$1\,lpm.  To explore the effect of oxygen content in the ambient gas, three gases were used at various times:  air (21\%~${\rm O}_2$), a 3\%~${\rm O}_2$/97\%~${\rm N}_2$ mix, and pure ${\rm N}_2$. The 3\%~${\rm O}_2$/97\%~${\rm N}_2$ mix was usually provided by premixed gas cylinders, but sometimes by mixing air with ${\rm N}_2$. The ${\rm N}_2$ came from compressed gas cylinders. Air was dried by flowing it through a ${\rm CaSO}_4$ desiccant column; gases from cylinders did not require desiccation.  During irradiation shutdowns, we measured the gas outflow oxygen content with an Apogee Instruments, model MO-200 oxygen meter. The gas coming out of the air and ${\rm N}_2$ boxes always had the correct oxygen concentration to within 0.1\%.  The 3\%~${\rm O}_2$ gas was always within 0.1\% of 3\%~${\rm O}_2$ when using premixed cylinders, and between  2\% and 5\%~${\rm O}_2$ when mixing air and ${\rm N}_2$.

The boxes were fed with gas through 1/8" copper tubing of sufficient length to equilibrate the gas temperature with the climate chamber temperature.
Positive flow was assured by bubblers (Amazon ASIN B00MV61HY4) on the boxes' gas outputs, filled with light viscosity mineral oil (McMaster-Carr \#3190K291), which also prevented air backflow.

Before each irradiation, to bring the oxygen concentration inside  each tile toward its equilibrium concentration for that temperature and atmosphere,
the gas mixtures were first flowed through the boxes at room temperature, then at the temperature of the irradiation.
 

 The boxes also contained \sipms\ that were not covered by a tile (``bare SiPMs") to monitor their irradiation damage.
The boxes had an input for an optical calibration signal.  
The boxes contained Pt1000 resistance temperature detectors (RTDs) (one close to each \sipm) used in 2-wire mode to monitor temperature.
Supply voltages and box output signals were carried via coaxial cables.  

Light from a 470\,nm  LED
outside the radiation area was used for the calibration signal.  The light was fed to an optical fanout box on a 15\,m quartz optical fiber (\#~FP600URT from Thorlabs, 600\,$\mu$m core diameter, high OH).
The fanout box had one input SMA bulkhead feedthrough that connected to a 600~$\mu$m fiber inside the box. An SMA junction inside the box connected the 600\,$\mu$m fiber to four 250 and one 100\,$\mu$m fibers, which were routed to SMA bulkhead output feedthroughs.
Outputs 
fed 2\,m \#~FP600URT light fibers to the scintillator boxes inside the radiation area, and  a 15\,m \#~FP600URT fiber to
a photodiode operating in photovoltaic mode
outside the radiation area. The photodiode was used to monitor the LED output.
The temperature at the location of the photodiode was stable enough that gain corrections were not needed.

The \sipm\ output voltages, RTD resistances, supply voltages, and the photodiode current were measured using a Keithley DAQ6510 and Keithley 7700 and 7702 multiplexer modules, which use mechanical switches.  Two Rigol DP821 power supplies provided the \sipm\ and LED supply voltages.
Digitization occurred 0.25\,s after switch closing to reduce noise.  The integration period was one power line cycle.
The Keithley DAQ and Rigol Power Supplies were controlled through SCPI commands from a GUI running on a laptop. All three devices were connected directly to the laptop over USB.
Because the breakdown voltage of \sipms\ is temperature dependent, the supply voltage was adjusted at different temperatures to maintain an overvoltage of 3--4~V.

\begin{figure}[hbtp]
\centering
\includegraphics[width=0.8\textwidth]{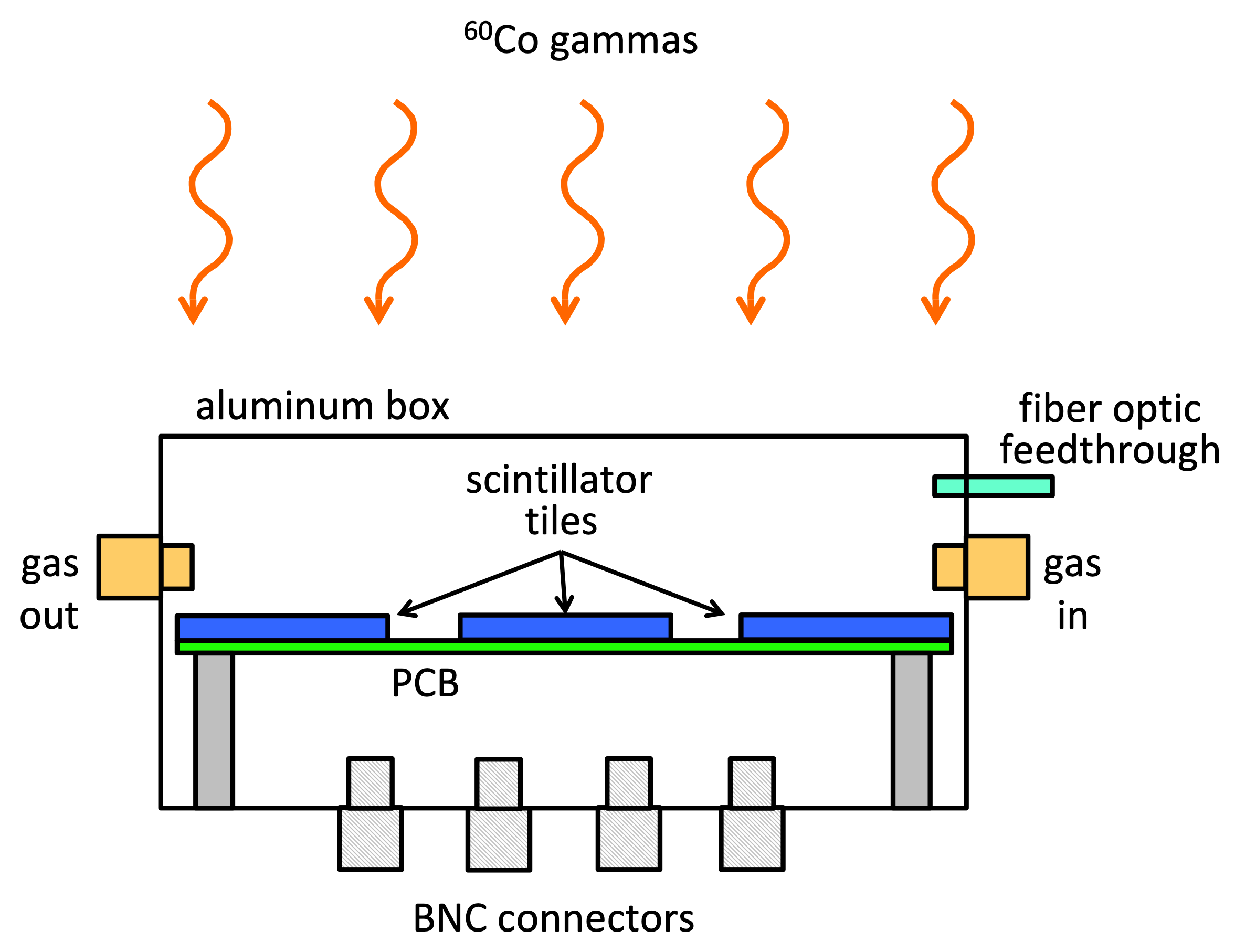}

\caption{
Schematic of the box used to hold the tiles;
}
\label{fig:boxSketch}
\end{figure}
\begin{figure}[hbtp]
\centering
\includegraphics[width=0.75\textwidth]{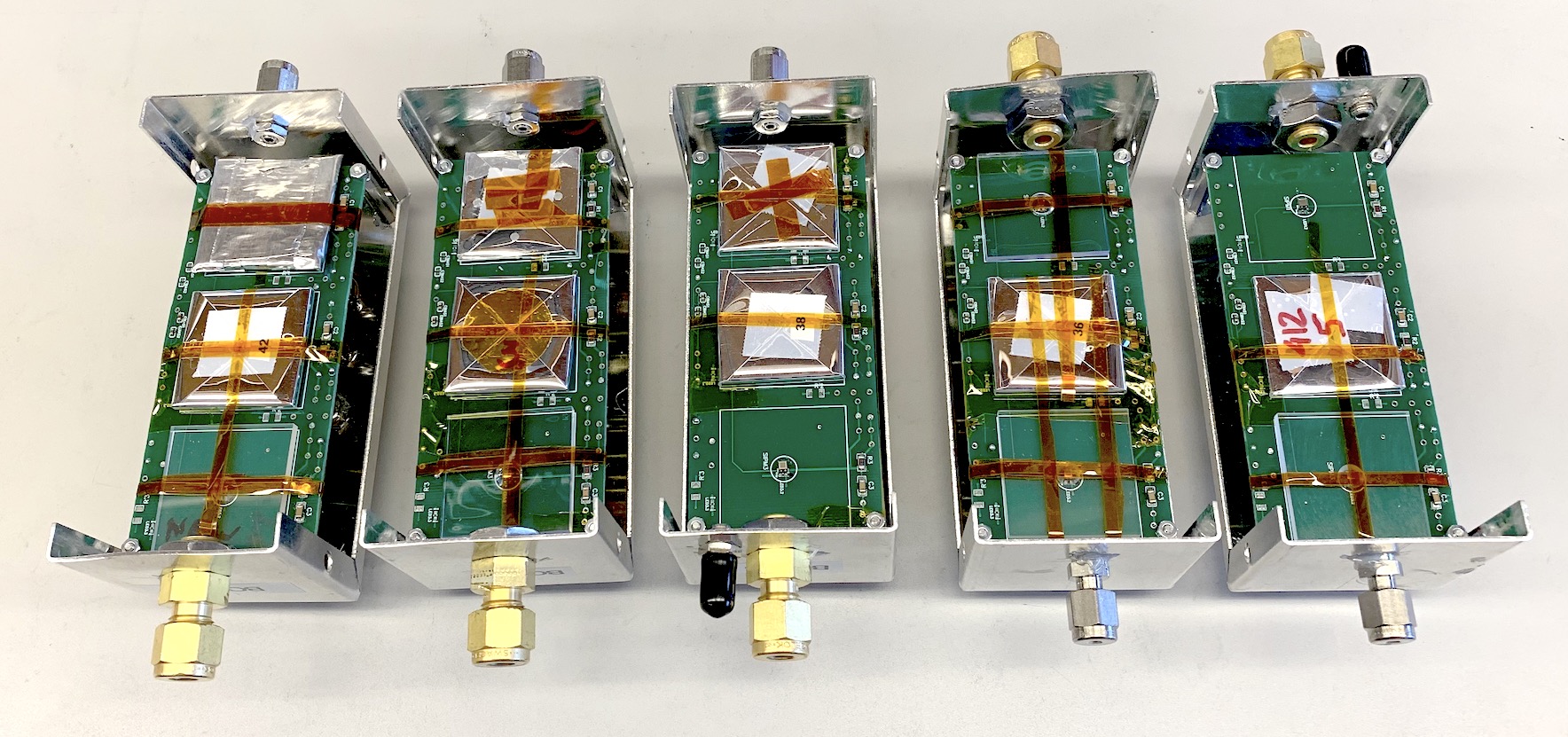}

\caption{
Photograph of five boxes with lids removed.
}
\label{fig:boxImage}
\end{figure}

Figure~\ref{fig:boxSketch} shows a schematic sketch of a box.
Figure~\ref{fig:boxImage} shows a photograph of five boxes with lids removed.

\begin{figure}[hbtp]
\centering
\includegraphics[width=0.65\textwidth]{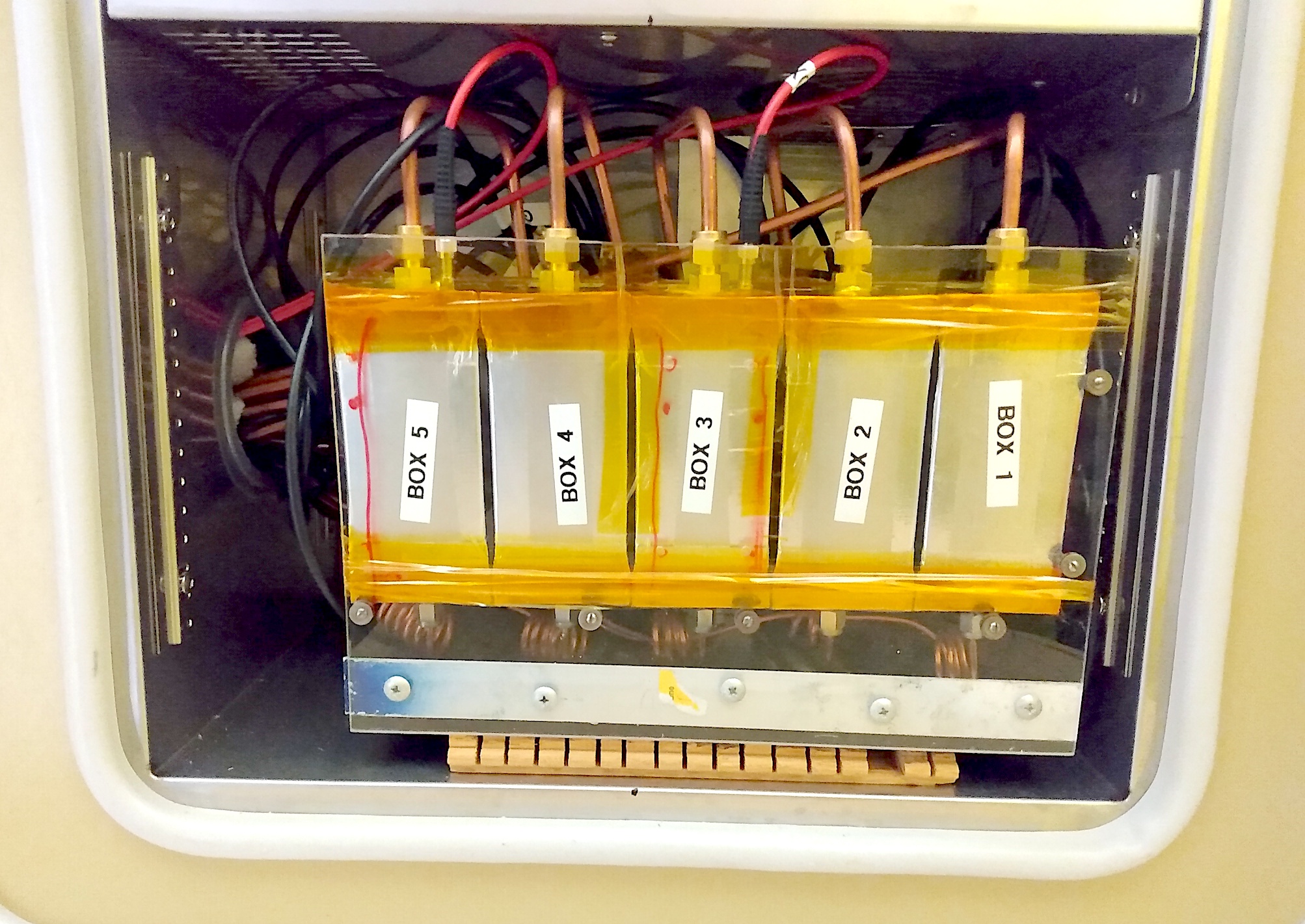}

\caption{
Photograph of the climate chamber (with door open) after being loaded with the boxes containing the tiles.
}
\label{fig:freeze}
\end{figure}

 The boxes were placed in a Tenney TJR-A-F4T climate chamber to control the temperature.
 Figure~\ref{fig:freeze} shows boxes installed in the climate chamber.

\section{Irradiation details}
\label{sec:irrad}
\begin{figure}[hbtp]
\centering
\includegraphics[width=0.65\textwidth]{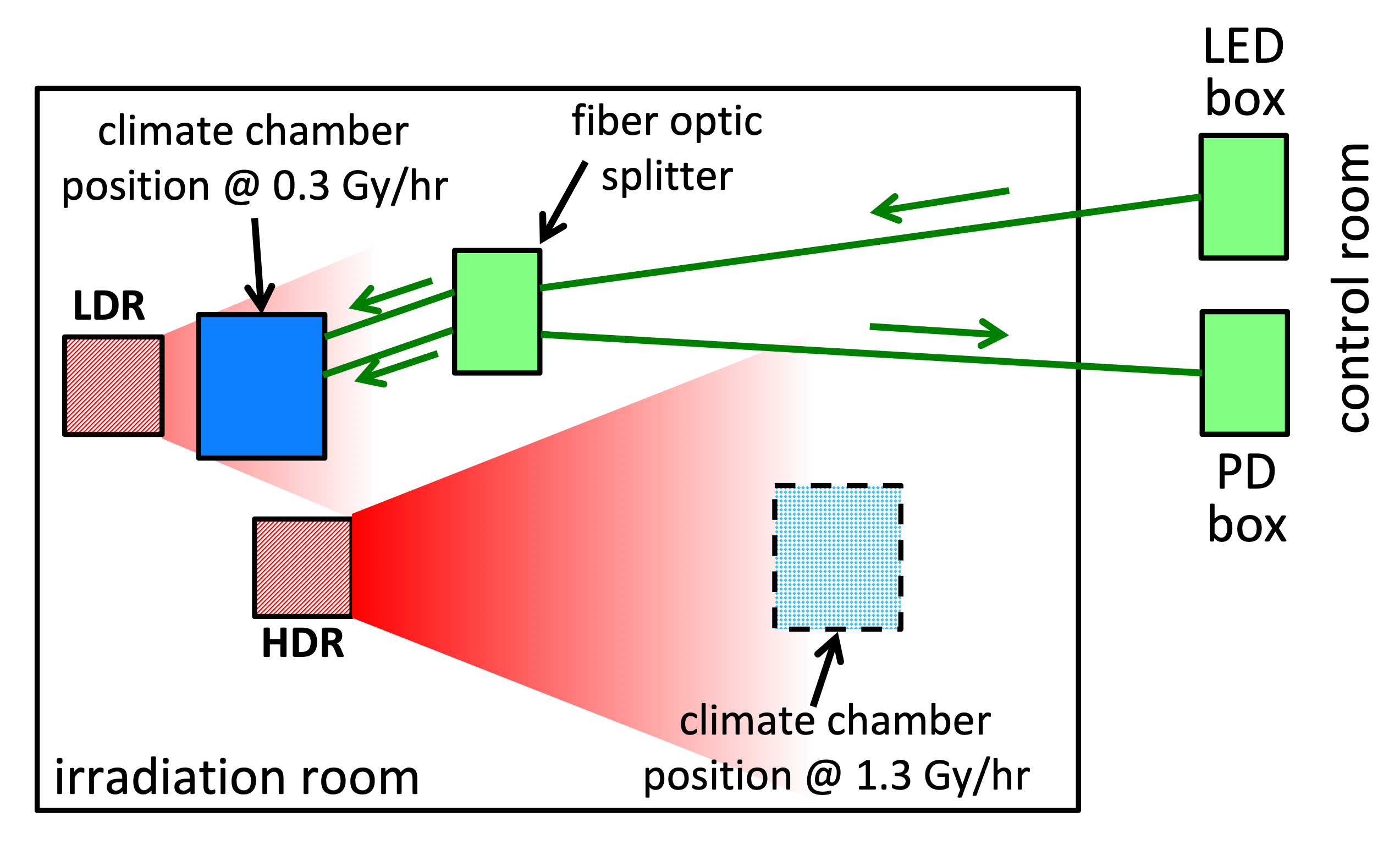}
\caption{
Schematic top view of the Goddard Space Flight Center \cosixty\ irradiation facility.  The labels ``LDR" and ``HDR" indicate the low-dose-rate  and the high-dose-rate irradiators, respectively.  Fiber optics are denoted by the green lines; the direction of light travel is indicated by green arrows.  The boxes holding the scintillator samples are located inside the climate chamber.  The sketch is drawn for the 0.3\,Gy/hr configuration, but the location of the climate chamber is also indicated for the 1.3\,Gy/hr and 1.6\,Gy/hr configuration.  The LED box contains the LED calibration light source.  The PD box contains a photodiode to quantify the light produced by the LED.
}
\label{fig:radarea1}
\end{figure}

Irradiations were done in two locations in the irradiation room at the \cosixty\ Radiation Effects Facility at Goddard Space Flight Center: one allowing the higher dose rates of \gdoseratewarm\,Gy/hr and \gdoserateh\,Gy/hr, the other the lower rate of \gdoseratel\,Gy/hr.  Dose rates have a 10\% systematic uncertainty.
Figure~\ref{fig:radarea1}  shows the experimental geometry in the radiation area.


Data presented in this paper come from three irradiation campaigns. The irradiations were  uninterrupted except for occasional shutdowns lasting  $\sim$1~hour each. The temperature and atmosphere were left unchanged during the shutdowns. 
An exception is the data plotted in Fig.~\ref{fig:r1}:  at about 20~Gy the irradiator was turned off for two days, and at about 40~Gy the irradiator was turned off for three days; the temperature was maintained at -30\degree\,C during these shutdowns.

Each campaign had different pre-irradiation gas flow durations. Most of the data presented here had gas flow at room temperature for 51 hours and at -30\degree\,C for 145 hours before the start of irradiation.  Figure~\ref{fig:r1} shows data taken after a pre-irradiation gas flow of 1.5~hours at -30\degree\,C.   The data in Fig.~\ref{fig:r2} [right] were taken in air at room temperature; this box had no gas purge flow.   The data for the upper two datasets in Fig.~\ref{fig:r6} [left] were taken after pre-irradiation flow of 71 hours at room temperature and 74 hours at -30\degree\,C.  Forty-eight hours into the annealing phase (Fig.~\ref{fig:r6}, [right]), the source of the 3\%~${\rm O}_2$ gas was switched from premixed cylinders to a mixture of air and ${\rm N}_2$ (2\% to 5\%~${\rm O}_2$).

\section{Calibration of the \sipms\ }
\label{sec:sipmcalib}

The light output of the tiles was measured as a voltage across the resistor in series with each tile's \sipm.  
This voltage has three components: the current related to the light incident on the \sipm, the \sipm\ dark current, and
a small pedestal (around 500\,$\mu$V at -30\degree\,C, equivalent to 10\,nA), whose behavior with temperature is consistent
with some component in the circuit acting as a thermocouple.  
The gain and photodetection efficiency (PDE) of the \sipm\ depend on its overvoltage.  
Its capacitance  depends on the voltage across the \sipm.
Because the \sipm\ was in series with the read out resistor, both the voltage across the \sipm\ (bias)  and the overvoltage depend on the \sipm\ current.
In this section, we describe how we correct the signal for changes in the system response due to these sources.


In order to measure the dark current and pedestal, the irradiator was periodically turned off for a few minutes during an irradiation, typically once per day.
For data taken at \gdoserateh\,Gy/hr, the average of these direct measurements was subtracted from the signal to yield the net signal from the incident light. The dark current and pedestal exhibited small fluctuations between measurements compared to the signal size. 
 In contrast, data taken at \gdoseratel\,Gy/hr and \gdoseratewarm\,Gy/hr were corrected using a linear interpolation between the closest dark current and pedestal measurements before and after the data. This interpolation resulted in an improved estimation for these measurements.

The dependence of the \sipm\ response on overvoltage was accounted for as follows. Either before or after a dark current plus pedestal measurement, a calibration curve for the \sipms\
was also measured. With the irradiator on, the bias was stepped from 2\,V under the nominal bias to 1\,V above, and then back down, in steps of 0.5\,V.
Measurements lasting several minutes were taken at each voltage. Data from these measurements, with its dark current and pedestal subtracted, were used to fit the \sipm\ response as a quadratic function of bias voltage at a fixed temperature $C(B)$. To correct the measured voltage at a measured bias $V(B)$ to what it would be at the nominal bias voltage $V(B_0)$, we use the ratio of the calibration curves at the two biases. The ratios

\begin{equation}
\frac{C(B_0)}{C(B)} = \frac{V(B_0)}{V(B)}
\label{eqn:correctionBias}
\end{equation}

\noindent are used to extract $V(B_0)$,  correcting all of the measured voltages to the nominal bias.

Two methods were used to correct for temperature variation over the course of the run. The first method, used by the \gdoserateh\,Gy/hr and \gdoseratewarm\,Gy/hr data, was effective for the small changes in temperature (typically less than a degree) experienced due to imperfect temperature regulation of the climate chamber. The \gdoseratel\,Gy/hr data used a different method 
effective for the large temperature change from -30\degree\,C to -5\degree\,C.

To include small temperature shifts in the first approach, Eq.~\ref{eqn:correctionBias} was adapted to include a temperature-dependent breakdown voltage. The largest temperature-based effect on \sipms\ is a linear shift in breakdown voltage with temperature, so accounting for the shift allows the overvoltage correction method to also correct for temperature. A linear correction of 37.6\,mV/\degree\,C to the breakdown voltage was used. In this correction method, $B$ from Eq.~\ref{eqn:correctionBias} was changed to $B-0.0376\Delta T$, where $\Delta T$ is the new temperature minus the current temperature. By decreasing $B$ as the temperature increases, we modeled the increased breakdown voltage at higher temperatures, and thus the shifted nominal bias.


Temperature shifts can also be modeled by using calibration curves in Eq.~\ref{eqn:correctionBias} taken at different temperatures. One curve is measured just prior to the temperature change and one after. If we have a measured voltage $V$ at bias $B$ and temperature $T$, we can correct it to the nominal bias and temperature using the ratios

\begin{equation}
\frac{C(B_0,T_0)}{C(B,T)} = \frac{V(B_0,T_0)}{V(B,T)}\,.
\label{eqn:correctionTemperature}
\end{equation}
As the curves are measured at two different times, the signal used to measure both calibration curves must be the same, and hence it cannot be used if the signal is annealing rapidly between the two measurements, or if the signal is being reduced through radiation damage. To ensure the stability of the signal this temperature correction was performed with data taken at the end of annealing when signals were changing very slowly. This correction can be used in conjunction with the linear correction for small temperature changes. We used this method to correct from -30\degree\,C to -5\degree\,C in the \gdoseratel\,Gy/hr data. The advantage of this method is that it includes all possible temperature effects on the response curve, including capacitance and tile light yield.

To check the stability of the calibration, the calibration curves were measured periodically. No dose or time dependence was seen. The continuity of corrected data between changes in calibration curve and bias subtraction indicate the validity of the method.

\section{Results}
\label{sec:results}

At the start of irradiation, the light
yield of a tile configuration depends on the brightness of its  scintillator,
the density of color centers created during the manufacturing process, and the ability of its wrapper to trap light.
Table~\ref{tab:initialyield} gives the measured \sipm\ 
current at the start of irradiation for the different tile configurations.
Tiles wrapped in ESR collect approximately twenty times more light than unwrapped tiles.

\begin{table}[!htp]
\centering
\caption{Current at start of irradiation for various
tile configurations. Unless otherwise noted starting temperature is -30\degree\,C.
For the ``holes" column, ``A" indicates holes while ``B" indicates slits.
}
\begin{tabular}{c c c c c c }
\hline
material & dose rate    & atmosphere, temperature & wrapping & holes & current \\
         & (Gy/hr) &            &          &       & ($\mu$A) \\
         \hline
BC-412 & \gdoseratel & 3\% ${\rm O_2}$ & ESR & - & 10.13 \\
BC-412 & \gdoseratel & 3\% ${\rm O_2}$ & ESR & - & 10.52 \\
BC-412 & \gdoseratel & ${\rm N_2}$ & ESR & - & 10.63 \\
BC-412 & \gdoseratel & Air & ESR & - & 7.49 \\
FNAL & \gdoseratel & 3\% ${\rm O_2}$ & ESR & - & 5.70 \\
BC-412 & \gdoseratel & 3\% ${\rm O_2}$ & ESR & A & 7.28 \\
BC-412 & \gdoseratel & 3\% ${\rm O_2}$ & ESR & B & 7.89 \\
BC-412 & \gdoseratel & 3\% ${\rm O_2}$ & Black & - & 0.45 \\
BC-412 & \gdoseratel & 3\% ${\rm O_2}$ & - & - & 0.44 \\
BC-412 & \gdoseratel &  ${\rm N_2}$ & - & - & 0.45 \\
BC-412 & \gdoseratel & Air & - & - & 0.40 \\
FNAL & \gdoseratel & 3\% ${\rm O_2}$ & - & - & 0.49 \\
FNAL & \gdoseratel &  ${\rm N_2}$ & - & - & 0.58 \\
\hline
BC-412 & \gdoserateh & 3\% ${\rm O_2}$ & ESR & -   & 38.14\\
BC-412 & \gdoserateh & Air & ESR & -               & 29.77\\
EJ-208 & \gdoserateh & 3\% ${\rm O_2}$ & ESR & -   & 38.94\\
EJ-208 & \gdoserateh & Air & ESR & -               & 30.79\\
BC-412 & \gdoserateh & Air & \tyvek & -            & 15.13\\
\hline
BC-412 & \gdoseratewarm & Air, 22\degree\,C & ESR & -      & 23.9\\
\hline
\end{tabular}
\label{tab:initialyield}
\end{table}

Because we acquired data from a given channel only once every several seconds, from these data we cannot rule out the possibility of a rapid ($\sim$seconds) initial decline in scintillator signals upon irradiation.  To test for a rapid initial decline, we employed a second, lower dose rate irradiator that contributed with a dose rate $10^{-3}$ that of the main irradiator, and thus would have caused minimal radiation damage. We measured \sipm\ signals using this low dose rate irradiator both before and after a one minute exposure to the primary irradiator (which was the first irradiation the tiles received).  A decrease in the ratio of after/before \sipm\ signals would indicate radiation damage in the first minute of irradiation.  We determined this ratio independently for tiles at -30\degree\,C in air and in 3\%~${\rm O}_2$.  For each gas, we computed the weighted average of the ratio for two similar tiles, EJ-208 and BC-412, both wrapped in  ESR.  We find the weighted mean for air is $1.003\pm0.004$, and for 3\%~${\rm O}_2$ is $0.999\pm0.004$; both are consistent with no damage from one minute of irradiation, so the values from Table~\ref{tab:initialyield} are accurate measurements of the initial signal.

\begin{figure}[hbtp]
\centering
\includegraphics[width=0.85\textwidth]{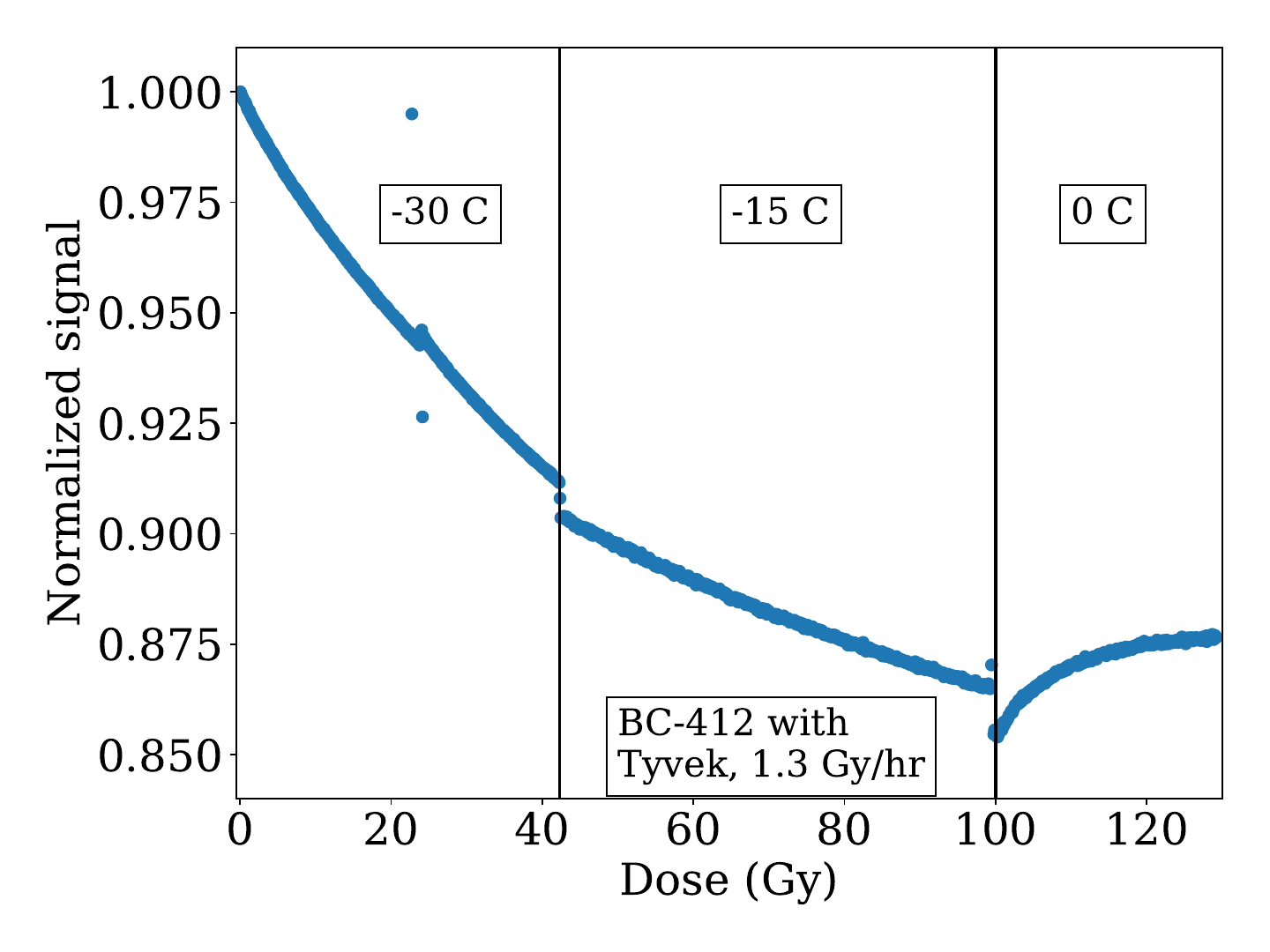}
\caption{
Ratio of the light yield during irradiation divided by the light yield at the start of irradiation for a BC-412 tile as a function of dose
and temperature.  The temperature at the start of the irradiation was  -30\degree\,C.  The temperature was changed at the dose corresponding to the first vertical line to  -15\degree\,C, and at the second line to 0\degree\,C. While the samples were at -30\degree\,C there were two long irradiation shutdowns, for about two days at 20~Gy and three days at 40~Gy.   This tile was irradiated at \gdoserateh\,Gy/hr in air. The tile was wrapped in \tyvek\ .
}
\label{fig:r1}
\end{figure}

Figure~\ref{fig:r1}  shows the effect of temperature on signal loss during irradiation, presumably due to increasing radical density. 
The dataset shows the relative light yield as a function of dose for a BC-412 tile wrapped in \tyvek\ in air.  
Initially the temperature was  -30\degree\,C.  
At the dose corresponding to the left vertical line, the temperature was changed to  -15\degree\,C. At the right vertical line,
the temperature was changed to 0\degree\,C.
During the first period at  -30\degree\,C, there is a rapid light loss.  When the temperature is changed to
-15\degree\,C, the slope changes and the rate of decrease of light loss slows.  When the temperature is changed to 
0\degree\,C, the light yield increases.  An increase is expected if the higher temperature allows sufficient motion of the radicals to significantly increase the probability of
annealing on the timescale of this measurement at this temperature.  At 0\degree\,C temperature and this dose rate, the light output becomes stable over the timescales probed with dose, indicating
that the rate of radical creation and annihilation are approximately equal.  The discontinuities at the temperature boundaries might be due to reduced scintillator light yield at higher temperatures \cite{9721821} or from imperfect temperature calibrations.

\begin{figure}[hbtp]
\centering
\includegraphics[width=0.45\textwidth]{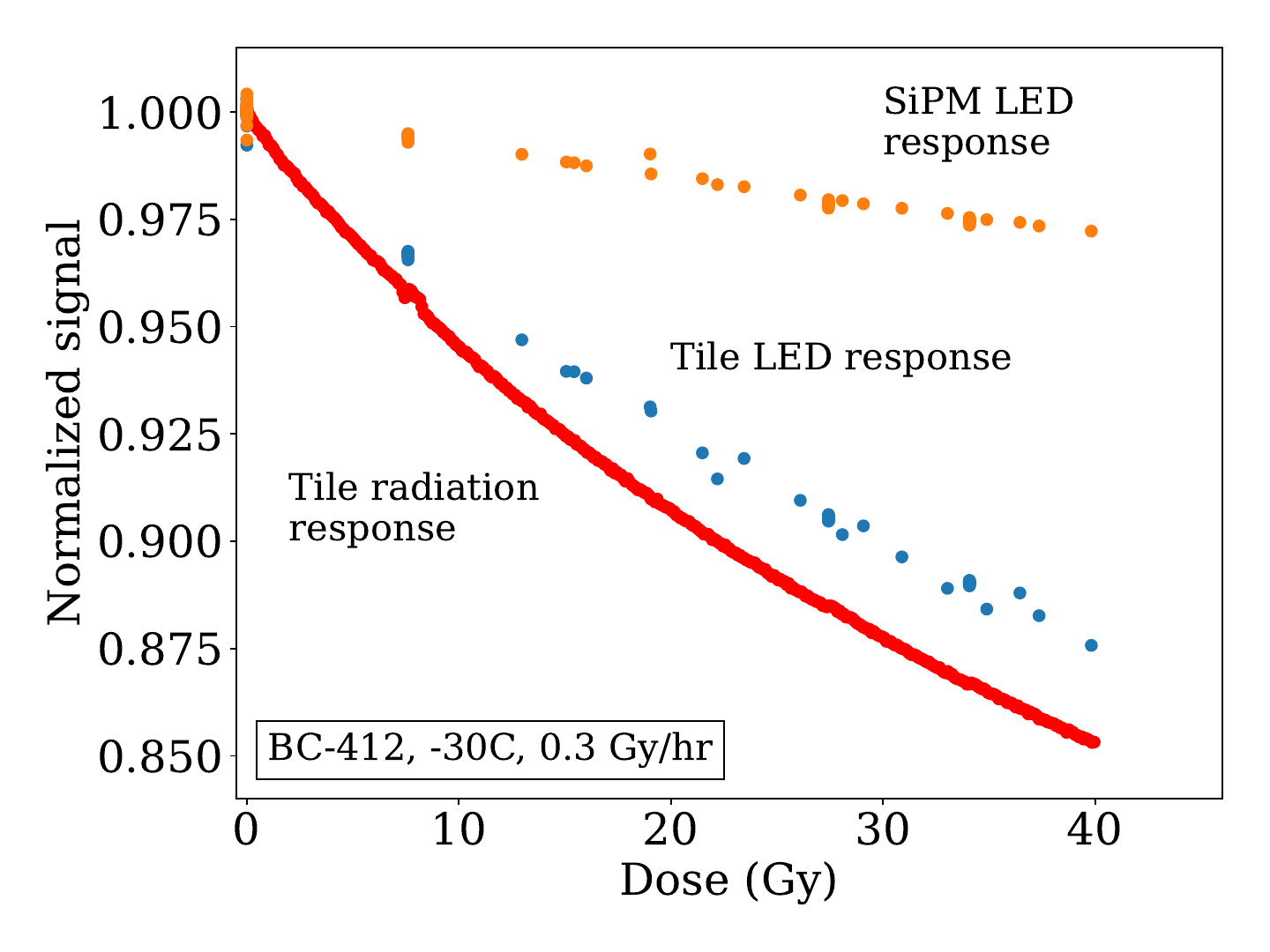}
\includegraphics[width=0.45\textwidth]{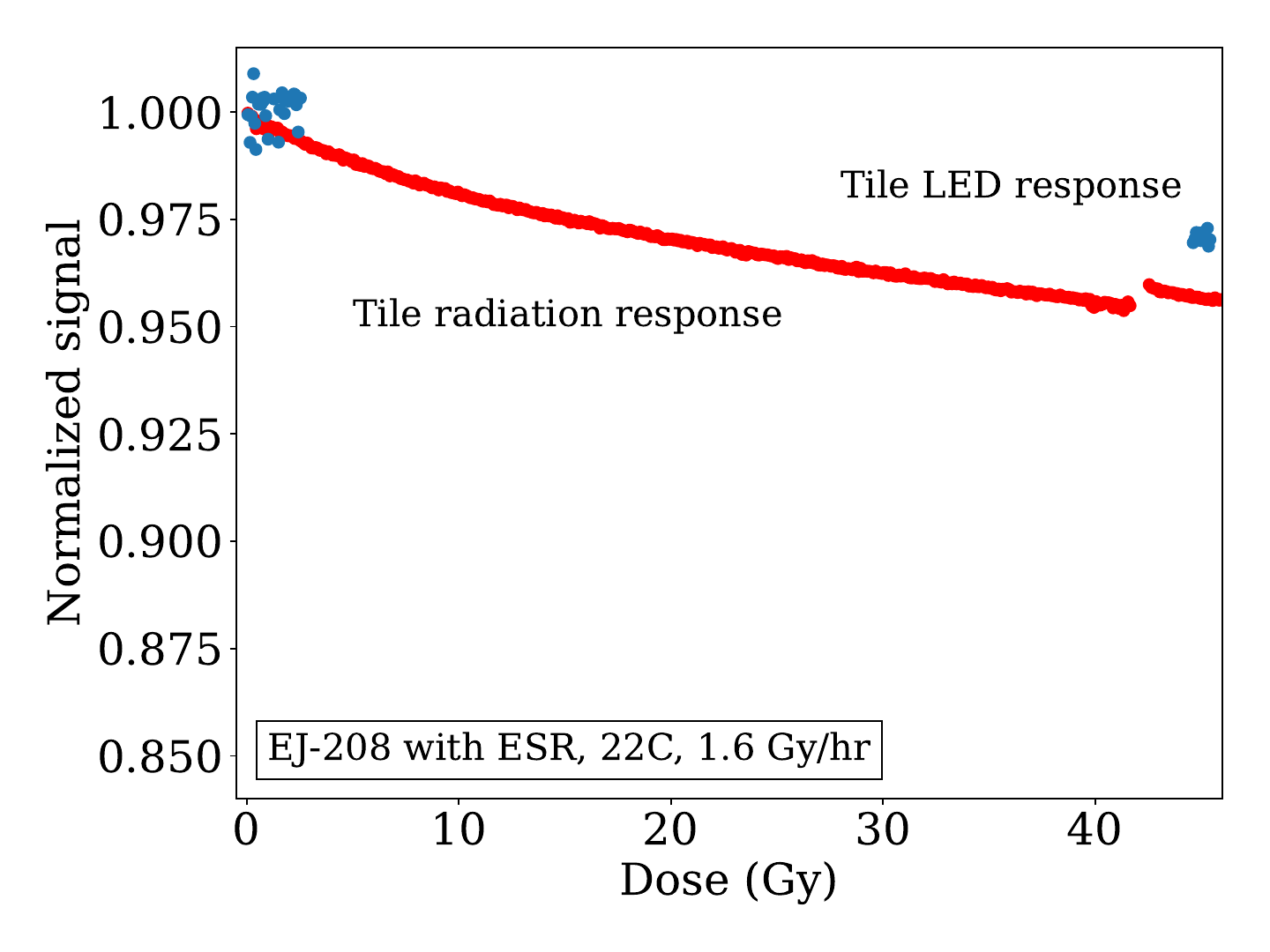}
\caption{
[left] Ratio of the light yield during irradiation divided by the light yield at the start of irradiation as a function of dose
for a BC-412 tile wrapped in ESR for a dose rate of \gdoseratel\,Gy/hr and a temperature of -30\degree\,C in 3\%~${\rm O_2}$.
The bottom dataset represents measurements using the light created by the \cosixty\ source while the middle represents the light that comes from the 
fiber from the LED, passes through the wrapper and tile, and illuminates the \sipm.  The signal from the ``bare" \sipms\ induced by the LED light
is also shown at the top.
[right] Ratio of the light yield during irradiation divided by the light yield at the start of irradiation as a function of dose
for a EJ-208 tile wrapped in ESR for a dose rate of \gdoseratewarm\,Gy/hr and a temperature of 22\degree\,C in air.
The bottom dataset represents measurements using the light created by the \cosixty\ source. The signal from the ``bare" \sipms\ induced by the LED light is shown at the top (blue). Due to possible radiation damage to the fiber optic cable affecting the LED normalization, the signal decrease for LED light could be lower by $2\%$ of the original signal at a dose of 40\,Gy \cite{DUMANOGLU2002444,RayburnT}.
}
\label{fig:r2}
\end{figure}

Figure~\ref{fig:r2} [left] shows the light yield as a function of dose divided by the light yield at the start of irradiation
for a BC-412 tile. The bottom dataset represents when the light is produced by the \cosixty\ source, while the middle curve represents when the \sipm\ measures light from the LED
passing through the wrapper and tile for 3\%~${\rm O_2}$ atmosphere at a temperature of -30\degree\,C and a dose rate of \gdoseratel\,Gy/hr.
Also shown is the damage to a ``bare" \sipm\ as measured using LED light (top dataset).  The \sipm\ damage during irradiation is visible, 
but still small compared to the signal. Radiation damage to the fiber optic cable could have affected the LED normalization. 
 If so,  the true signal loss for LED light measurements could be lower by $2\%$ at 40\,G ~\cite{DUMANOGLU2002444,RayburnT} as drops of this size are seen in the light collected at the monitor photodiode during the run. 
 For the tile, the observed loss during the irradiation is large compared to the predictions from Eq.~\ref{eqn:oldprediction} for permanent damage.
The effective dose constant is estimated by fitting to a double exponential 

\begin{equation}
 \frac{L(d)}{L_0}=A_1\exp{\left(-\frac{d}{D_1}\right)} + A_2\exp{\left(-\frac{d}{D_2}\right)\,.}
    \label{eqn:doubleexp}
\end{equation}
A double exponential was chosen as it fit the data well, while a single exponential did not.

Results from the fit for the irradiated tiles (including this one) are given in Table~\ref{tab:fits}. Residuals of the fits had a standard deviation of about $0.05\%$ of the signal.

Figure~\ref{fig:r2} [right] shows results for a similar tile irradiated at room temperature.  The
damage is much smaller at room temperature, and is only slightly larger than the damage to the \sipm.
The damage to the \sipm\ is similar at both temperatures.
A possible explanation for the temperature dependence of scintillator damage is that the mobility of the radicals allows faster annealing at room temperature.
Another possibility is that oxygen permeates and binds more efficiently to the radicals at room temperature. The cause of the discontinuity in the bottom dataset at about 40\,Gy in not known.

\begin{table}[!htp]
\centering
\caption{Results of fits to light yield ratio as a function of dose to the functional form given in Eq.~\ref{eqn:doubleexp}. The formula line at the bottom of each dose rate section gives the expected values from Eq.~\ref{eqn:doseconst}. Uncertainties in the contribution to the signal
drop from \sipm\ damage is $\le$2\% and the uncertainties in the dose and dose rate are 10\%. For the ``wrapping \& holes" column, ``A" indicates holes while ``B" indicates slits.}
\begin{tabular}{c c c c c c c c }
\hline
material & dose rate   & atmosphere & wrapping  & $A_1$ & $A_2$ & $D_1$ & $D_2$  \\
         & (Gy/hr) &            &  \& holes  &       &       &(kGy)&(Gy) \\
         \hline
BC-412 & \gdoseratel & 3\% ${\rm O_2}$ & ESR   & 0.92 & 0.08 & 0.46 & 17 \\
BC-412 & \gdoseratel & 3\% ${\rm O_2}$ & ESR   & 0.92 & 0.08 & 0.49 & 18\\
BC-412 & \gdoseratel & ${\rm N_2}$ & ESR       & 0.92 & 0.08 & 0.53 & 18\\
BC-412 & \gdoseratel & Air & ESR               & 0.94 & 0.07 & 0.50 & 17\\
FNAL & \gdoseratel & 3\% ${\rm O_2}$ & ESR    & 0.93 & 0.07 & 0.72 & 14\\
BC-412 & \gdoseratel & 3\% ${\rm O_2}$ & ESR, A    & 0.94 & 0.07 & 0.56 & 19\\
BC-412 & \gdoseratel & 3\% ${\rm O_2}$ & ESR, B    & 0.94 & 0.06 & 0.54 & 19\\
BC-412 & \gdoseratel & 3\% ${\rm O_2}$ & Black  & 0.99 & 0.01 & 1.4 & 16\\
BC-412 & \gdoseratel & 3\% ${\rm O_2}$ & -     & 0.99 & 0.01 & 1.3 & 14\\
BC-412 & \gdoseratel &  ${\rm N_2}$ & -         & 0.99 & 0.01 & 2.0  & 14\\
BC-412 & \gdoseratel & Air & -             & 0.99 & 0.01 & 1.1 & 13\\
FNAL & \gdoseratel & 3\% ${\rm O_2}$ & -       & 0.98 & 0.02 & 1.4 & 14\\
FNAL & \gdoseratel &  ${\rm N_2}$ & -           & 0.98 & 0.02 & 1.5 & 12\\
Formula & \gdoseratel &  - & -                 & 1.00 &   -  & 6.2 & -\\
\hline
BC-412 & \gdoserateh & 3\% ${\rm O_2}$ & ESR  &  0.88 & 0.12 & 0.67 & 27 \\
BC-412 & \gdoserateh & Air & ESR                &  0.90 & 0.11 & 0.70 & 27 \\
EJ-208 & \gdoserateh & 3\% ${\rm O_2}$ & ESR   &  0.87 & 0.13 & 0.74 & 30 \\
EJ-208 & \gdoserateh & Air & ESR                &  0.90 & 0.10 & 0.71 & 28 \\
BC-412 & \gdoserateh & Air & \tyvek             &  0.98 & 0.02 & 0.55 & 8.7 \\
Formula & \gdoserateh &  - & -                 & 1.00 &   -   &  13   & -\\
\hline
BC-412 & \gdoseratewarm & Air, 22\degree C & ESR       & 0.97 & 0.04 & 3.6 &  16\\
Formula & \gdoseratewarm &  - & -                 & 1.00 &   -      & 14 & -\\
\hline
\end{tabular}
\label{tab:fits}
\end{table}

\begin{figure}[hbtp]
\centering
\includegraphics[width=0.45\textwidth]{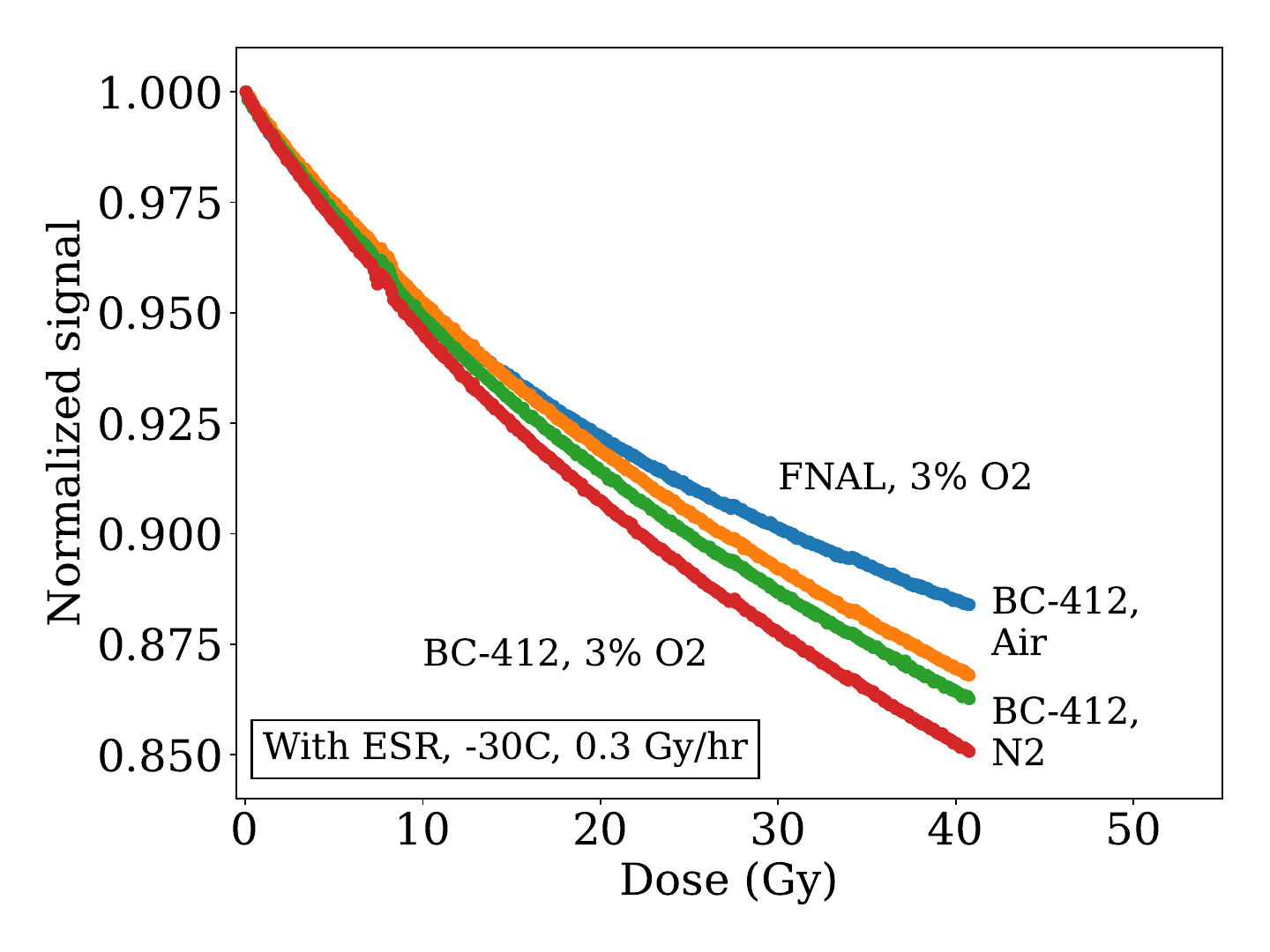}
\includegraphics[width=0.45\textwidth]{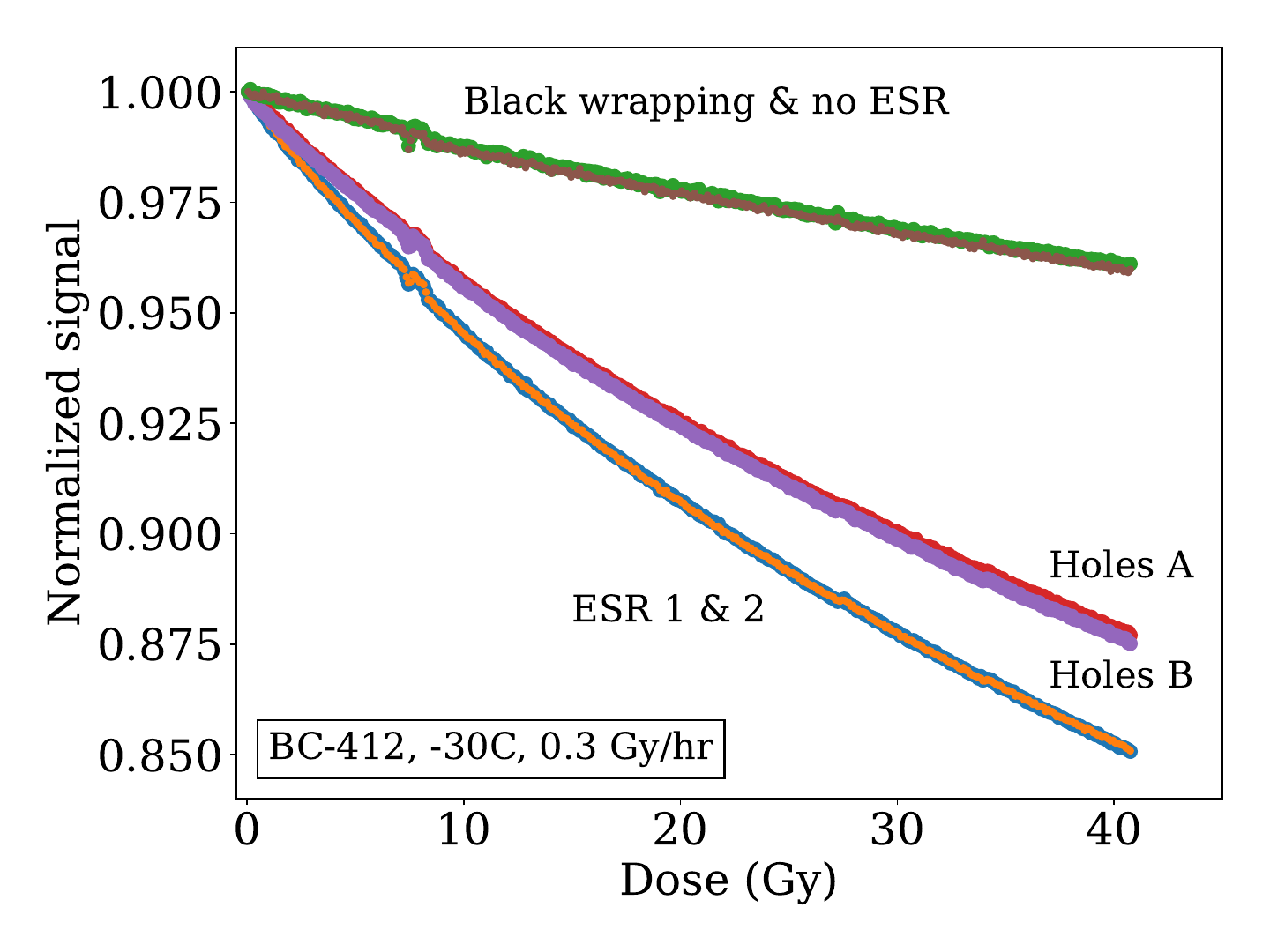}
\caption{
[left] Ratio of the light yield during irradiation divided by the light yield at the start of irradiation as a function of dose
for various atmospheres for the BC-412 and FNAL tiles with ESR wrappings. All four tiles were irradiated at -30\degree\,C with a dose rate of 0.3 Gy/hr. The top dataset represents the FNAL tile in 3\% oxygen, the top middle curve BC-412 in air, the bottom middle BC-412 in nitrogen, and the bottom curve BC-412 in 3\% oxygen.
[right] Ratio of the light yield during irradiation divided by the light yield at the start of irradiation as a function of dose
for various wrappings of BC-412 tiles. All six tiles were irradiated at -30\degree\,C in 3\% oxygen with a dose rate of 0.3 Gy/hr. The bottom overlapping datasets represent tiles wrapped in ESR, the upper middle is ESR with holes, the bottom middle is ESR with slits, and top two overlapping are black optical paper and no wrapping. 
}
\label{fig:r4}
\end{figure}

Figure~\ref{fig:r4} [left] compares the relative signal loss of cast PVT-based Bicron tiles at -30\degree\,C and  a dose rate of 0.3\,Gy/hr
for three different oxygen concentrations: standard air, 3\% oxygen, and pure nitrogen.  Also shown is a PS-based FNAL tile in 3\% oxygen.
The molded PS tiles show less damage than the cast. It is not known if this is due to the difference in material or to the shaping procedure.
The behavior with oxygen concentration is not understood and warrants further study.

Figure~\ref{fig:r4} [right] examines the effects of wrapping material on the relative light loss.  The more reflective the coating,
the greater the measured signal, but also the greater the relative light loss.  This is consistent with formation of color centers: the more reflective coatings allow greater light collection by trapping photons until they bounce into the \sipm, while the light detected in unwrapped tiles will tend to have a more direct path.  The longer paths will suffer more absorption from color centers, and a larger fraction of these photons will be removed.  This model is also consistent with the measurements of the  tile wrapped in black paper, which also suppresses photons with long path lengths.  Since the radiation damage datasets in Fig.~\ref{fig:r4} [right] are identical for black wrapping and no wrapping, the pathlength model is favored.  The black paper was wrapped in aluminum tape to make the wrapping sealed hermetically (except for the dimple hole); if oxygen in the atmosphere is important for reducing radiation damage, the damage could be more severe in the black-wrapped tile than in an unwrapped tile, which is not observed. This indicates that either the dimple hole provides significant access to oxygen to the entire tile, or oxygen is playing a minimal role in the observed radiation damage.

\begin{figure}[hbtp]
\centering
\includegraphics[width=0.45\textwidth]{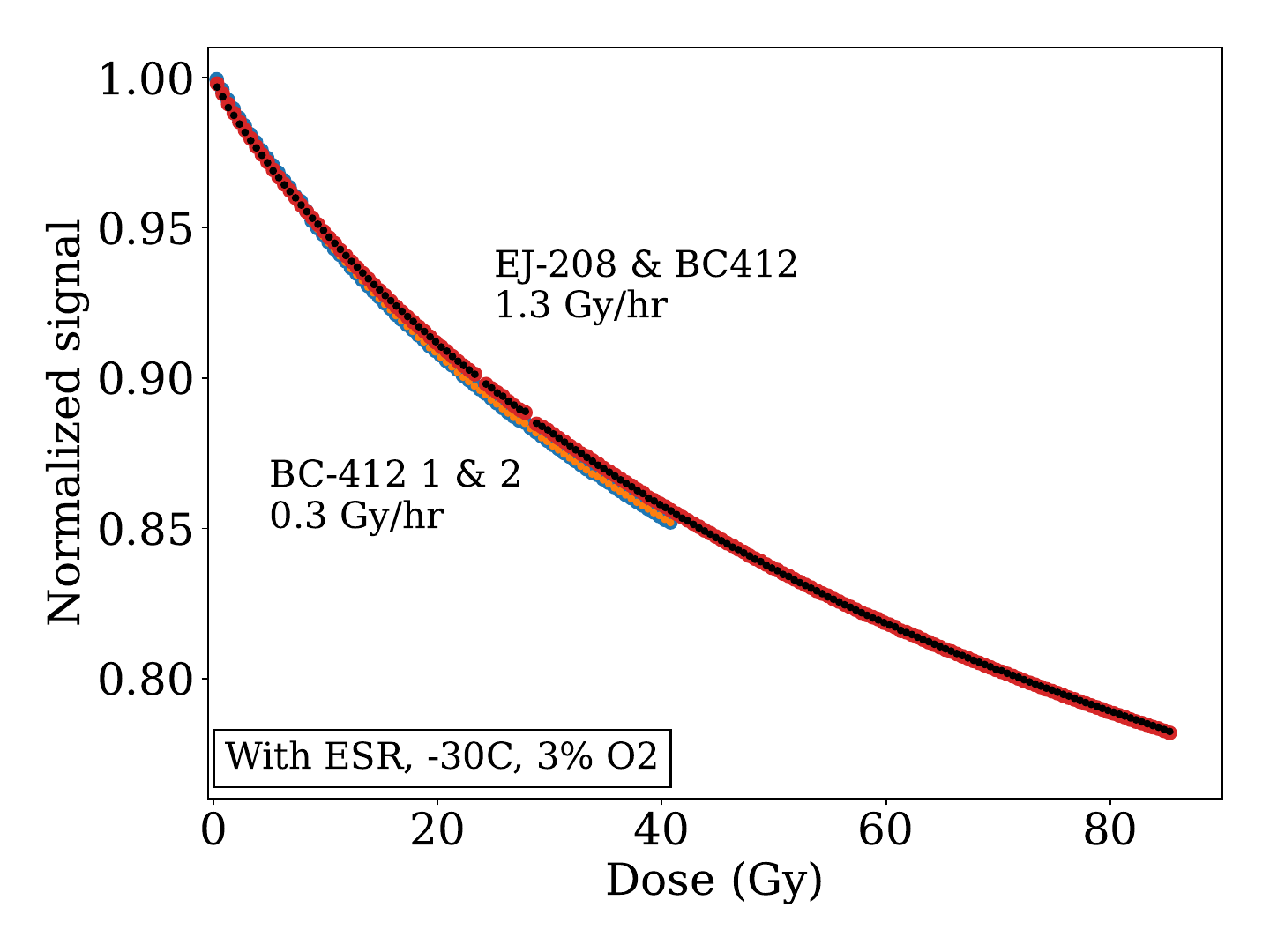}
\includegraphics[width=0.45\textwidth]{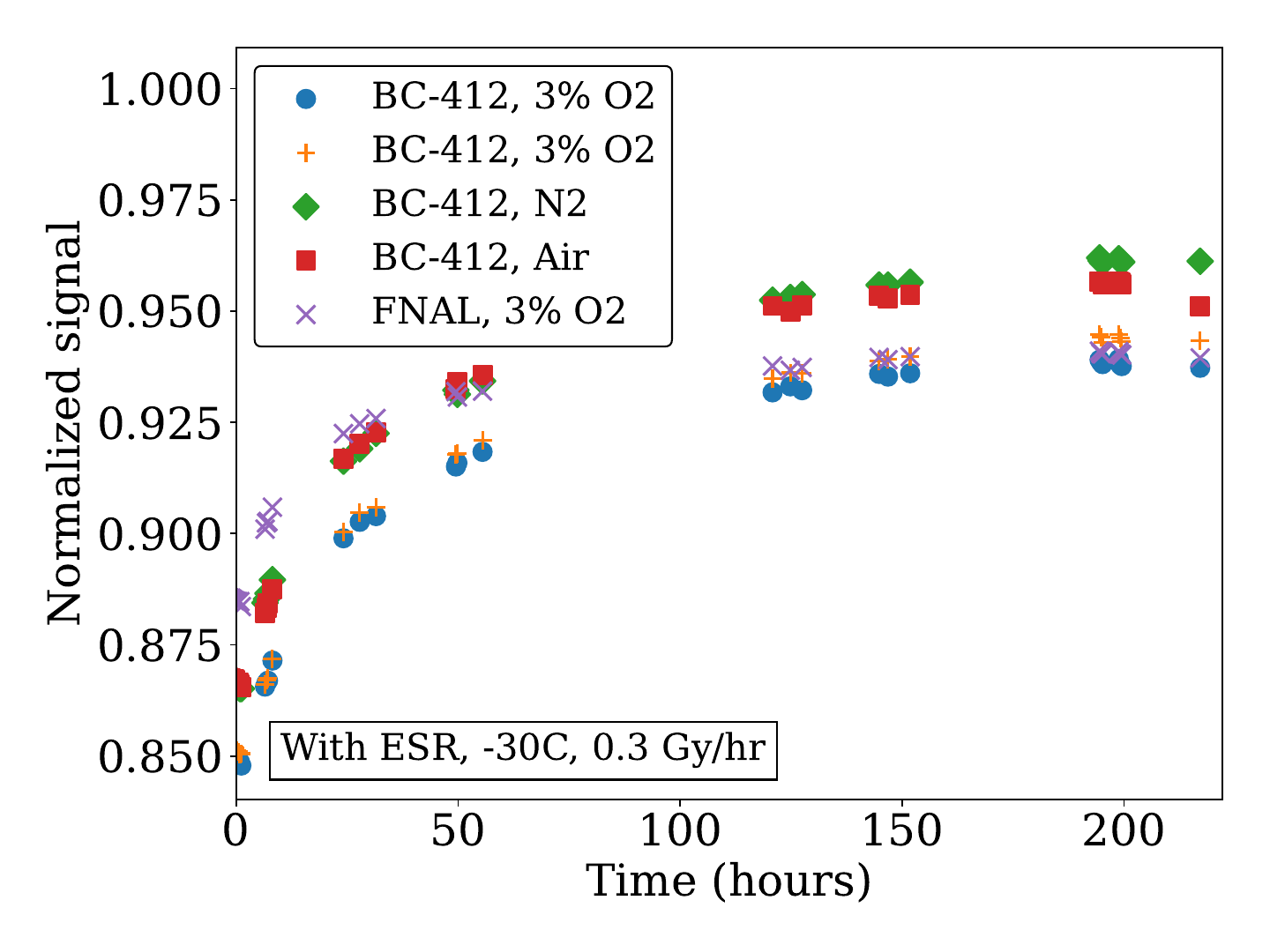}
\caption{
[left]
Ratio of the measured light yield divided by the light yield at the start of irradiation as a function of dose. The bottom overlapping datasets were irradiated at 0.3\,Gy/hr while the top overlapping datasets were irradiated at \gdoserateh\,Gy/hr. One of the top datasets was EJ-208, while the other three were BC-412. All tiles were irradiated at -30\degree\,C in 3\% oxygen and were wrapped in ESR.
[right]
Ratio of the measured light yield divided by the light yield at the start of irradiation as a function of dose, showing the annealing of the tiles wrapped in ESR. The end of irradiation is t = 0. while the square marked tile was irradiated in 21\% oxygen, and the remaining tiles were irradiated with 3\% oxygen.
 }
\label{fig:r6}
\end{figure}

Figure~\ref{fig:r6} [left] compares measurements of light loss at the two different dose rates.  
No dose rate dependence is seen.

Figure~\ref{fig:r6} [right] shows the annealing of the temporary damage after irradiation for tiles irradiated at -30\degree\,C wrapped in ESR for various oxygen environments and for both molded PS-based FNAL and cast PVT-based BC-412 tiles.  The annealing time scale is similar for all combinations.

\section{Conclusions}
\label{sec:conclusion}
The first measurements of temporary damage to plastic scintillators during irradiation by a \cosixty\ source
at  dose rates of \gdoseratel, \gdoserateh, and \gdoseratewarm\,Gy/hr were presented.  
Evidence of substantial temporary damage at temperatures of -30\degree\,C and -15\degree\,C was seen.  The data taken at -30\degree\,C was well fit by a double exponential and showed no sign of a plateau in the radical density up to
the achieved dose of 40\,Gy.  
At 0\degree\,C, a plateau at about 90\% of the initial light output was seen.
No dose rate dependence was seen at -30\degree\,C.  
Because the relative light loss was greater for ESR wrapped tiles than bare tiles, the loss mechanism was more
consistent with color center formation than initial light loss.
The behavior with oxygen
concentrations between 0\% and standard air was surprising, in that the change with oxygen content was not monotonic.  Future investigation is warranted.

These results should be useful to experiments planning to use plastic scintillator at cold temperatures in a radiation environment.  Further study, however, is needed to understand in more detail the dependence of the light reduction on the surrounding atmosphere and for large doses.

\section{Acknowledgments}
The authors would like to thank the staff at Goddard Space Flight Center's Radiation Effects Facility for being a delight to work with during the irradiations. 
We would like to thank Felix Sefkow, Jeremy Mans, Ted Kolberg, Mitch Wayne and  Iouri Musienko for useful conversations and suggestions. We would like to thank the HGCAL collaboration for their assistance and resources.
 This work was supported in part by U.S. Department of Energy Grant DESC0010072.

\bibliography{main}

\end{document}